\documentclass[aps,preprint, amsmath,amssymb,author-numerical,superscriptaddress,nofootinbib]{revtex4}
\usepackage{graphicx,color,dcolumn,subfigure,bm}

\begin{document}
\def\De{{\rm De}}

\title{Fluid elasticity increases the locomotion of flexible swimmers}

\author{Julian Espinosa-Garcia}
\affiliation{Instituto de Investigaciones en Materiales, Universidad Nacional Aut\'onoma de M\'exico, M\'exico D.F. 04510, M\'exico.}
\author{Eric Lauga}
\email{elauga@ucsd.edu}
\affiliation{Department of Mechanical and Aerospace Engineering, University of California, San Diego, 9500 Gilman Drive, La Jolla, California 92093-0411, USA.}
\author{Roberto Zenit}
\email{zenit@unam.mx}
\affiliation{Instituto de Investigaciones en Materiales, Universidad Nacional Aut\'onoma de M\'exico, M\'exico D.F. 04510, M\'exico.}

\date{\today}
\begin{abstract}

We conduct experiments with flexible swimmers to address the impact of fluid viscoelasticity on their locomotion. The swimmers are composed of a  magnetic head actuated in rotation by a frequency-controlled magnetic field and a flexible tail whose deformation leads to forward propulsion.
We consider both viscous Newtonian and  glucose-based Boger fluids with similar viscosities.  We find that the elasticity of the fluid systematically enhances the locomotion speed of the swimmer, and that this enhancement increases with Deborah number.  Using Particle Image Velocimetry to visualize the flow field,  we find  a significant difference in the amount of shear  between the rear and leading parts of the swimmer head. We conjecture that viscoelastic normal stresses lead to a net elastic forces in the swimming direction and thus a faster swimming speed.
\end{abstract}

\maketitle

A field originally started by Sir G. I. Taylor \cite{taylor51}, the fluid dynamics of swimming microorganisms  was very active in the 1970's \cite{jahn72, lighthill75, brennen77,purcell1977,childress81}. Renewed interest was recently  prompted by new series of questions arising on nonlinear and nonlocal swimming behavior, including cell-cell hydrodynamic interactions, collective locomotion,  instabilities of active suspensions, and synthetic swimming systems \cite{lp09}. One of such questions, and the focus of this paper, concerns locomotion in complex fluids.

Many situations of biological importance involve locomotion in, and transport of, polymeric fluids with large relaxation times in viscous dominated regimes. Well-studied examples include mucus transport by lung cilia  \cite{sleigh88} and locomotion of mammalian spermatozoa in cervical mucus \cite{suarez06}.  Except for a handful of investigations \cite{ross1974,chaudhury1979,fulford1998}  all past experimental and theoretical work in the field focused solely on  Newtonian fluids.

Recently,  a series of studies has addressed locomotion in  complex fluids \cite{lauga07,FuPowersWolgemuth2007,FuWolgemuthPowers2008,FuWolgemuthPowers2009,lauga_life,teran2010,arratia2011,Liu2011,Harman2012}, with somewhat contradictory results. Analytical studies in the  small-deformation limit  showed that in a Oldroyd   viscoelastic fluid, locomotion under a given gait (two- or three-dimensional waving) always lead to a decrease of the swimming speed with the Deborah number, {\rm De}, to ratio of the fluid relaxation time scale to the typical time scale of the swimming motion  \cite{lauga07,FuPowersWolgemuth2007,FuWolgemuthPowers2009,lauga_life}. In contrast, numerical simulations for a two-dimensional swimmer in an Oldroyd fluid   showed that, for large-amplitude motion, an increase in swimming speed could be obtained, which the authors attributed to high-strain regions behind the swimmer  \cite{teran2010}.

Similar to the modeling approaches, experimental investigations have lead to two different results. Shen and  Arratia studied the locomotion of {\it C. elegans} nematodes in shear-thinning polymeric fluids and found, in agreement with the small-deformation analysis, that viscoelasticity hinders locomotion \cite{arratia2011}.  Liu {\it et al.} \cite{Liu2011} considered the force-free translation of externally-rotated thin rigid helices, as a model for the dynamic of bacterial flagella in complex fluids. They found that, for large-amplitude motion in a constant-viscosity Boger fluid \cite{boger1977}, the velocity leading to the force-free condition  (equivalent to a swimming speed) could be increased by the presence of fluid elasticity.

In this paper we address the effect of fluid elasticity on flexible swimmers. We measure the swimming speed of magnetic bodies equipped with flexible tails actuated externally by a time-periodic magnetic field. The combination of filament flexibility, periodic actuation, and drag forces from the fluid break the time-reversibility for the filament shape and leads to forward propulsion \cite{WigginsGoldstein}. In a Boger fluid with constant viscosity we show that fluid elasticity always enhance the locomotion speed of the swimmers. In addition, the ratio between the non-Newtonian velocity and that measured in a Newtonian flow  systematically increases with the Deborah number.

\begin{figure}[b]
{\includegraphics[width=0.65\textwidth]{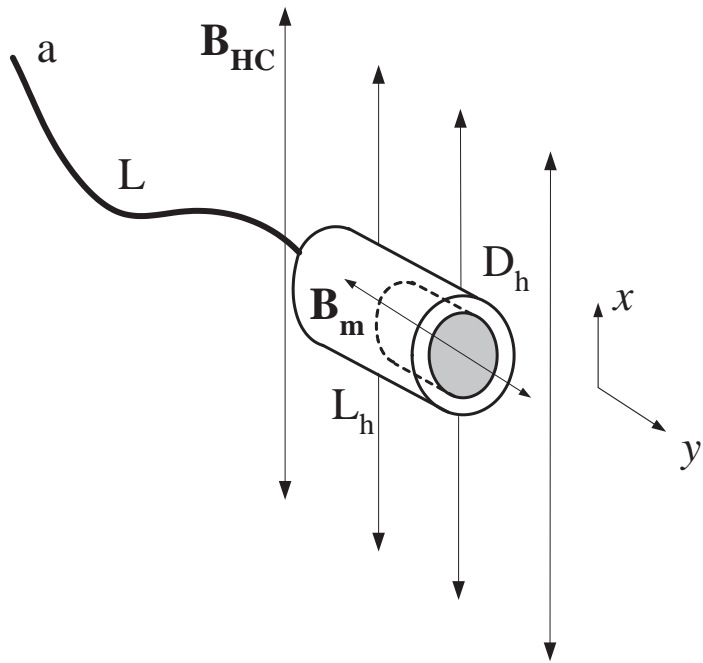}}
  \caption{Sketch of the magnetic swimmer with flexible tail. See text for notation.}
  \label{fig:setup}
\end{figure}

A sketch of the swimmer is shown in Fig. \ref{fig:setup}. The permanent magnet  is a rare Earth rod magnet (Magcraft, model NSN0658) with lengths and diameters of 3.18 mm possessing a remanent magnetic flux of $B_m=1.265\pm 0.015$ T.  Since the density of the magnet is larger than that of the test fluids ($\rho_{m}= 7450$ kg/m$^3$), it is encapsulated in a piece of plastic tubing in which an air bubble is  captured to make it neutrally buoyant; the end of the tubing is sealed with silicone rubber. The diameter and length of the head are $D_h=5.2$ mm and $L_h= 14.2$ mm. The tail,  glued to the head, is a piece of optic fiber of length $L=25$ mm  and radius  $a=62.5$ $\mu$m, with a Young's modulus similar to that of glass, $E\approx80 $ GPa.
The video \cite{Video1} shows a typical experiment (swimmer in a Boger fluid). Two additional swimmers are also built. One had a rigid tail, made from of hypodermic tubing, ($L=25$ mm, $a=125$ $\mu$m, $E=200$ GPa) and is used to demonstrate that without tail flexibility no swimming can occur \cite{purcell1977,{Video2}}. The second swimmer has a flat rubber tail ($L=20.5$ mm, thickness $t=0.8$ mm, width $w=3.4$ mm, $E= 0.01$ GPa) and is used  for measurements of the fluid velocity around the swimmer using a standard two-dimensional particle image velocimetry system (PIV, Dantec Dynamics).

\begin{figure}[t]
\includegraphics[width=0.65\textwidth]{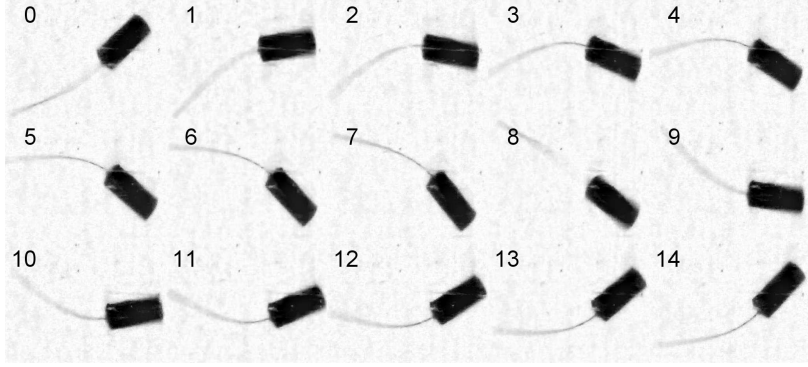}
  \caption{(Image sequence of a swimming cycle at frequency $\omega= 1$ Hz in the Boger fluid (B2);  time step between each image, $\Delta t$= 66.6 ms.}\label{fig:setup2}
\end{figure}

The external magnetic field, $B_{HC}(t)$, which changes direction in time, is generated by a Helmholtz coil pair \cite{kraftmakher2007}, with a radius of $R=140 $ mm and approximately $230$ wire turns in each coil. For a maximum current of 4 A, a magnetic field of about $B_{HC}|_{max}= 6$ mT is achieved. The coil pair is energized using a power supply that applies a DC voltage changing sign periodically, conferring the change in direction to the magnetic field, with a frequency of up to 10 Hz. The swimmers are placed inside a cylindrical container (diameter, $D_c=50.8$ mm). This  container is  located inside the Helmholtz coil which provides the driving magnetic actuation to rotate the magnetic head leading to deformation of the tail and locomotion. The motion of the swimmers was filmed with a digital camera, as illustrated in Fig. \ref{fig:setup2} over one  period of oscillation. The obtained images were processed digitally to obtain the position, speed, and inclination angle of the head in time, as illustrated in Fig \ref{fig:setup2}.

\begin{figure}[t]
\includegraphics[width=0.65\textwidth]{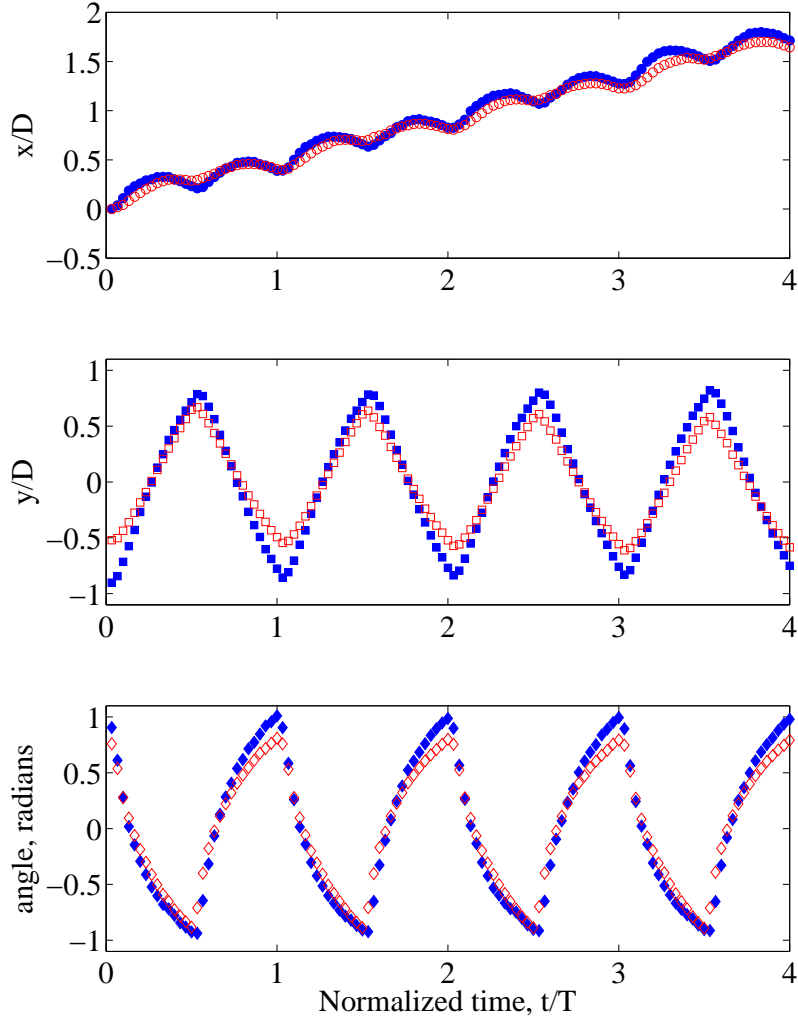}
  \caption{(Color online) Position of the head along ($\circ$) and across ($\Box$) the swimming direction as well as head angle ($\diamond$); filled  symbols (blue online) show results for the Boger (B2) fluid while empty symbols (red online) are for the Newtonian (N2) fluid.}\label{fig:setup2}
\end{figure}

\begin{table}[t]
  \centering
  \begin{tabular}{c|c|c|c|c|c}
  Fluid & G/W/PaaM& $\rho$& $\mu$ & n & $\lambda$\\
   & (\%) & (kg/m$^3$) & (Pa.s) &  & (s)\\
\hline
  Newtonian (N1) & 89/11/0 & 1390 & 3.5 & 1.0 & 0.0 \\
  Boger   (B1) & 84.96/15/0.04 & 1340 & 3.7 & 0.98 & 1.23 \\
  Newtonian (N2) & 89/11/0 & 1400 & 2.8 & 1.0 & 0.0 \\
  Boger   (B2) & 84.96/15/0.04 & 1350 & 2.7 & 0.98 & 1.03 \\
\hline
\end{tabular}
  \caption{Properties of the fluids studied: composition, density, viscosity, power index, and mean relaxation time. }\label{table:fluid_properties}
\end{table}

Close attention was paid to the preparation of the test fluids,  water solutions of industrial grade glucose. Two types of fluids are prepared,  Newtonian (N1 and N2) and viscoelastic Boger (B1 and B2). To confer elasticity to the glucose solutions, a small amount of polyacrylamide (Separan AP30, Dow Chemicals) is added and solutions with nearly constant viscosity and finite relaxation times are obtained \cite{boger1977}. In Table \ref{table:fluid_properties} we show the properties of the all fluids used; note that fluids N1 and B1 are more viscous than N2 and B2.  The density, $\rho$,  is measured using a pycnometer (Simax, 50 ml).
The rheological properties are characterized using a rheometer with a cone-plate geometry (TA Instruments, AT 1000N). Both steady and oscillatory tests are conducted to measure the shear viscosity, $\mu$, and the storage and loss moduli, $G'(\omega)$ and $G''(\omega)$, respectively. Once the Boger fluids are characterized, the Newtonian fluids are prepared by varying the amount of water to match the value of the shear viscosity of the viscoelastic fluids. For both fluids,  the maximum  Reynolds number obtained experimentally,  $Re=U D_h \rho/\mu$, where $U$ is the mean swimming speed, is less than 10$^{-3}$. The power index for the viscoelastic fluids,  $n$, is very close to one, hence their shear viscosity is nearly independent of shear rate. The viscoelastic effects quantified by our study are therefore attributable only to the elasticity of the fluids and not to their shear-dependance. The mean relaxation times are calculated considering the scheme proposed in  Ref.~\cite{Liu2011}. The experimental values of $G'(\omega)$ and $G''(\omega)$ are fitted to a generalized Maxwell model \cite{baumgaertel1989}, $G' (\omega)= \sum_{i=1}^{N}({g_i \lambda_i^2 \omega^2})/({1+\lambda_i^2 \omega^2})$  and
$G'' (\omega)=\omega \mu + \sum_{i=1}^{N} ({g_i \lambda_i \omega})/({1+\lambda_i^2 \omega^2}) $, with $N=4$ leading to an excellent fit. The values of the fitting parameters ($g_i$ and $\lambda_i$) are then used to estimate the mean  relaxation time as $\lambda=({\sum_{i=1}^{N} g_i \lambda_i^2})/({\sum_{i=1}^{N}g_i \lambda_i})$. Note that the value of the solvent viscosity was 3.4 and 2.6 Pa s for the fluids B1 and B2, respectively.

In Fig.~\ref{fig:setup2} we show a typical measurement of the swimmer kinematics in our study: position of the head along ($\circ$) and across ($\square$) the swimming direction as well as head angle ($\diamond$). Filled  symbols (blue online) correspond to the  Boger (B2) fluid while empty symbols (red online) are for the Newtonian (N2) fluid. As the swimmer head moves its head sideways, a net forward motion is produced. Even though the exact same swimmer was used in the two fluids under the same  magnetic driving, the motion in the viscoelastic case appears to be larger,  both for  sideways and forward displacements. The angle amplitude of the head appears to  also be larger in the Boger fluid.

\begin{figure}
    \includegraphics[width=0.65\textwidth]{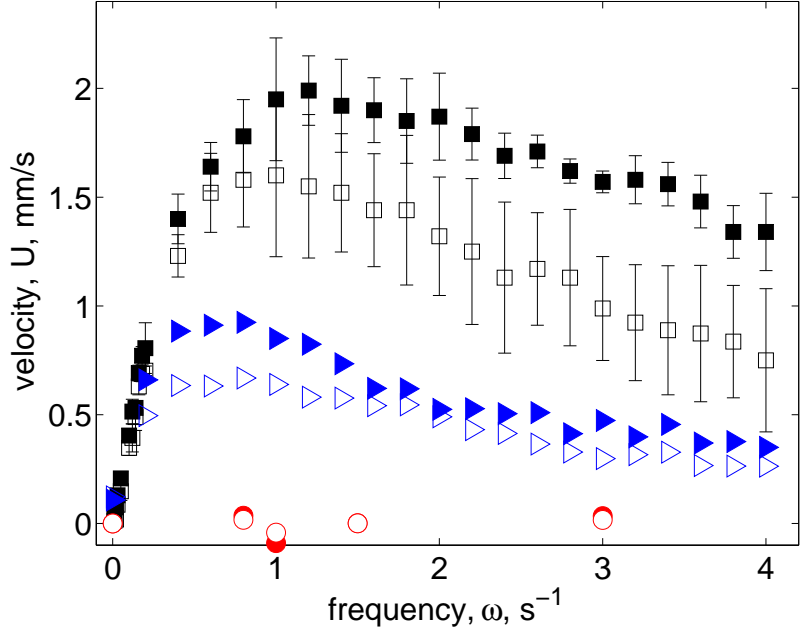}
  \caption{(Color online) Mean forward swimming speed, $U$ (mm/s), as a function of the oscillating frequency, $\omega$ (Hz). Filled and empty symbols correspond to Boger and Newtonian fluids, respectively. $\square$ (black online): fluids N2 and B2, flexible swimmer; $\triangleright$ (blue online): fluids N1 and B1,  flexible swimmer; $\circ$ (red online):  fluids N1 and B1, rigid tail swimmer. The errorbars depict the standard deviation of the measurements for the N2 and B2 fluids    which were repeated four times.}\label{fig:speed}
\end{figure}

In Fig.~\ref{fig:speed} the measured mean forward velocity, $U$ (mm/s), is shown as a function of the oscillating frequency, $\omega$ (1/s),
for the flexible and rigid tails. The results with the flexible flat tail are similar to the ones with a flexible cylindrical tail and are omitted for clarity. In all fluids, bodies with a rigid tail cannot swim. With flexible tails, as the oscillation frequency increases from zero, the forward velocity increases,  reaches a maximum and  decreases slowly. This is similar to past measurements for Newtonian fluids \cite{yu06}. In all cases, the swimming speed is seen to be larger for the viscoelastic fluids than that reached in the Newtonian fluid.  Our results are therefore in qualitative agreement with the numerical results of \cite{teran2010} and the experiments of \cite{Liu2011}  showing enhanced swimming velocity in viscoelastic fluids.

\begin{figure}
    \includegraphics[width=0.65\textwidth]{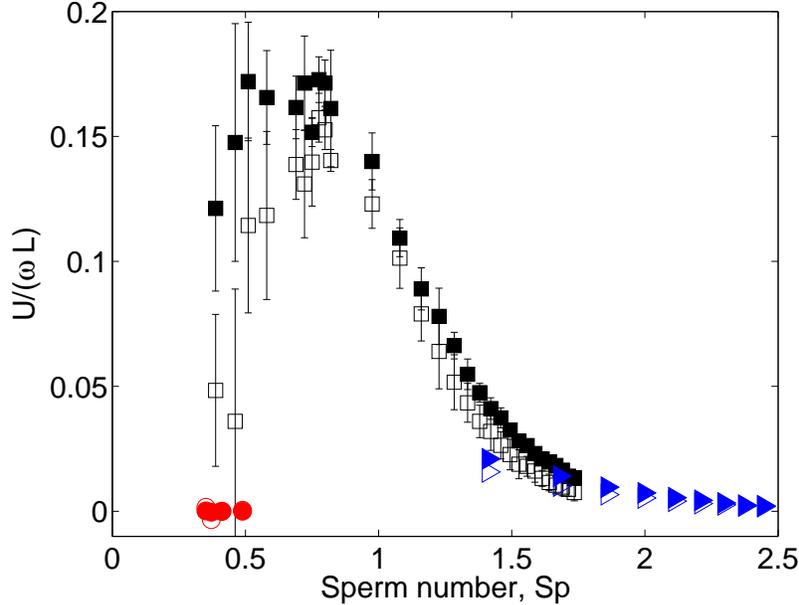}\\
  \caption{(Color online)  Normalized swimming speed, $U/\omega L$, as a function of sperm number, $\rm Sp$. All symbols as in Fig. \ref{fig:all_results}.}\label{fig:all_results}
\end{figure}

To help rationalize our results we plot our measurements in dimensionless terms. The relative importance of the bending forces in the tail and viscous drag forces are quantified by the so-called sperm number  \cite{lp09}
\begin{equation}\label{sperm_number}
    {\rm Sp}=L \left(\frac{\omega \varepsilon_\bot}{E I}\right)^{1/4},
\end{equation}
where $\omega$ is the oscillation frequency, $\varepsilon_\bot$ is a viscous resistance coefficient for fluid motion perpendicular to the tail, $E$ is the material's Young's modulus, and $I$ is the second moment of inertia of the tail cross-sectional area ($I=\pi a^4/4$ for a circular cross section). With the approximate value
$\varepsilon_\bot \approx  {4\pi \mu}/{\ln (L/a)}$, the sperm number can be written as
\begin{equation}\label{sperm_number_2}
        {\rm Sp}=2 \left(\frac{\mu \omega}{E}\right)^{1/4} \frac{L^*}{(\ln L^*)^{1/4}},
\end{equation}
where $L^*=L/a$ is the dimensionless tail length. The quantity in the first parenthesis on the right-hand side of Eq.~\eqref{sperm_number_2}, $\mu \omega/E $, is a pseudo-Weissenberg number that compares the viscous stress in the fluid with the elastic stress in the tail. In Fig.~\ref{fig:all_results} we plot  the dimensionless swimming velocity, $U/(\omega L)$,  as a function of $\rm Sp$.
Maximum swimming occurs at a sperm number of order one. For viscoelastic fluids, the swimming velocities are larger than in the Newtonian case  but the dependence with $\rm Sp$ appears to be similar.

\begin{figure}[b]
    \includegraphics[width=0.65\textwidth]{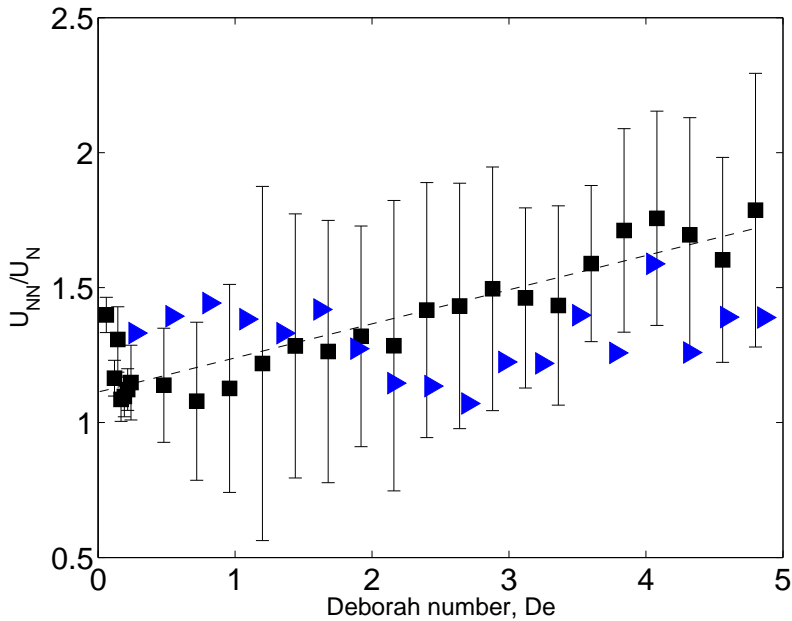}
  \caption{Ratio of non-Newtonian ($U_{NN}$) to Newtonian ($U_{N}$) swimming speeds, $U_{NN}/U_N$,  as a function of the Deborah number, $\rm De$.
  $\blacksquare$ (black online): fluids N2 and B2; $\blacktriangleright$ (blue online): fluids N1 and B1.   The dashed line is a linear fit to the data for all fluids.}\label{fig:Deborah}
\end{figure}

To quantify the differences in locomotion between the two types of fluids, we show in Fig.~\ref{fig:Deborah}  the ratio of the non-Newtonian ($U_{NN}$) to the Newtonian ($U_{N}$) swimming speeds, $U_{NN}/U_N$,  as a function of Deborah number,  ${\rm De}=\lambda \omega$. Despite the uncertainties due to variations in our measurements, the trend is clear (the line is a linear fit to the data for all fluids). Swimming is always faster in a non-Newtonian fluid, and the velocity ratio between the two fluids systematically  increases with Deborah number. The other two investigations which showed an increase of swimming in complex fluids \cite{teran2010,Liu2011} have found that there is a critical Deborah number at which the velocity increase is maximum. In contrast here, and for the range of parameters tested, we observe a continuous increase of the swimming velocity for De numbers as large as 5.

\begin{figure}[t]
{\mbox{\subfigure[]{\includegraphics[width=0.5\textwidth]{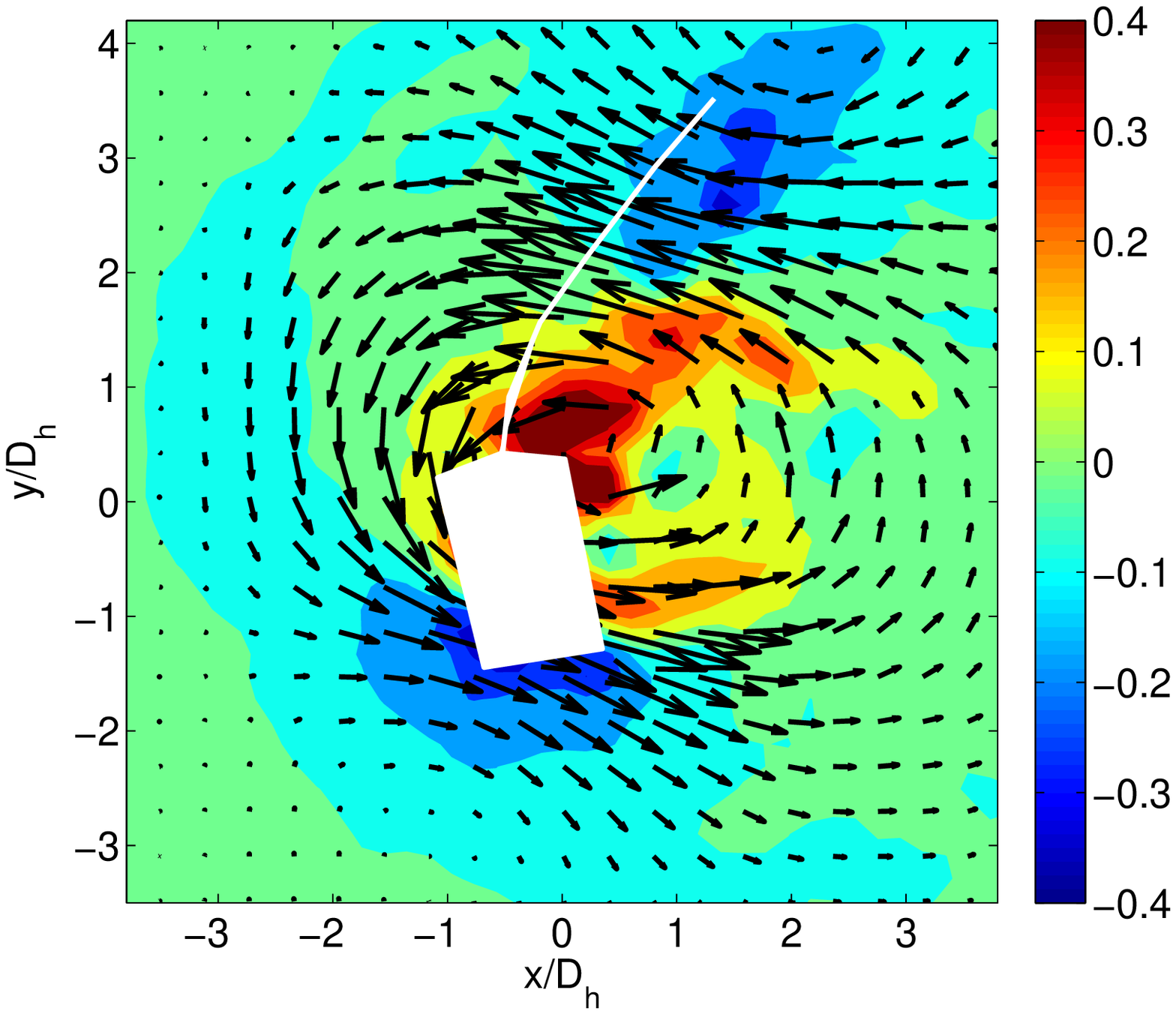}}}}
{\mbox{\subfigure[]{\includegraphics[width=0.5\textwidth]{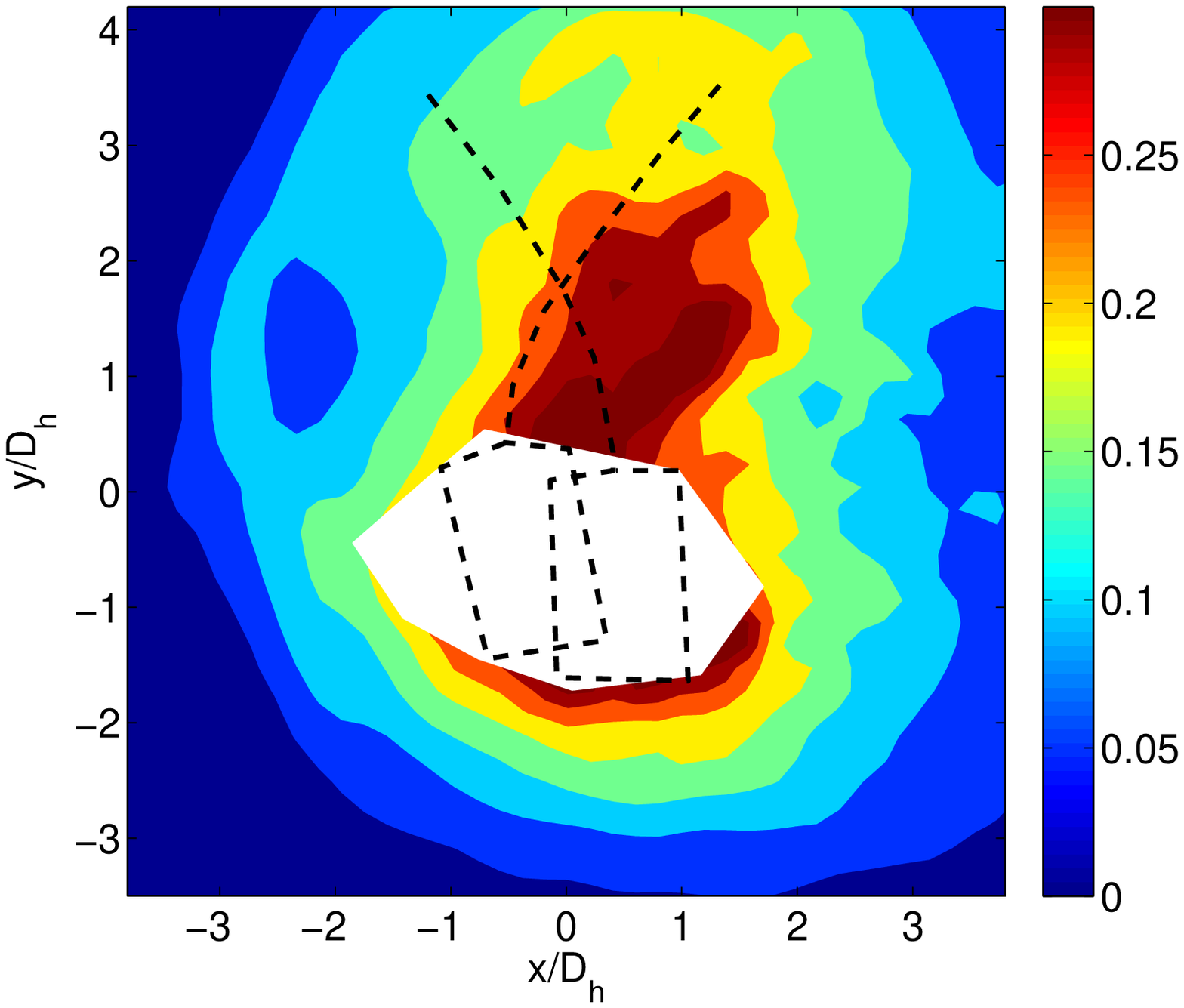}}}}
  \caption{(Color online) PIV results for the flow field around the swimmer with a two-dimensional tail in fluid B2. (a): Vorticity (colors) and velocity (arrows)  fields; (b): Cycle-averaged magnitude, $|\textbf{D}|$, of the rate-of-strain tensor, $\textbf{D}$.  Vorticity and rate-of-strain fields are normalized by the oscillation frequency, $\omega$. The white regions show the position of the swimmer at different instants throughout the cycle.}\label{fig:D}
\end{figure}

Why does this swimmer move systematically faster in a viscoelastic media? To help rationalize our results, we conducted a visualization study of the velocity fields around the swimmer\cite{Video3}. We employed  two-dimensional PIV  to characterize the velocity field around the swimmer with a flat tail. In Fig.~\ref{fig:D}a we show the typical velocity and vorticity fields around the swimmer at a given instant in the beating period. A large vortical structure develops around the swimmer; during each cycle of  oscillation, a vortex forms, dissipates quickly and a new one forms  in the opposite direction.

With our measurement of the velocity field, we can compute all four in-plane components of the velocity gradient tensor. The rate-of-strain tensor, $\textbf{D}$, can therefore be measured around the swimmer, $\textbf{D}=\frac{1}{2}[\nabla \mathbf{v}+(\nabla \mathbf{v})^T]$, where $\mathbf{v}$ is the velocity in the fluid. We show in Fig.~\ref{fig:D}b  the field of the phase-averaged magnitude of $\textbf{D}$, defined as $|\textbf{D}|=\langle\sqrt{\textbf{D}:\textbf{D}}\rangle$, where  $\langle \cdot \rangle$ denotes averaging in time. This plot quantifies therefore  regions where shear deformations are large throughout the periodic beating of the swimmer tail. Clearly, on average, the rear part of the head of the swimmer is subject to a larger shear than its front.

As is well known,  viscoelastic fluids  subject to shear deformations  result in additional normal stresses, due to the stretching of polymeric molecules along flow streamlines  \cite{barnes1999}.  Based on our observation of the difference in the  amount of shear between the leading and trailing sides of the head, we conjecture that it is the presence of these non-Newtonian normal stresses which induce a nonzero elastic force in the swimming direction, leading to an increase in the swimming speed.

\begin{figure}
\includegraphics[width=0.65\textwidth]{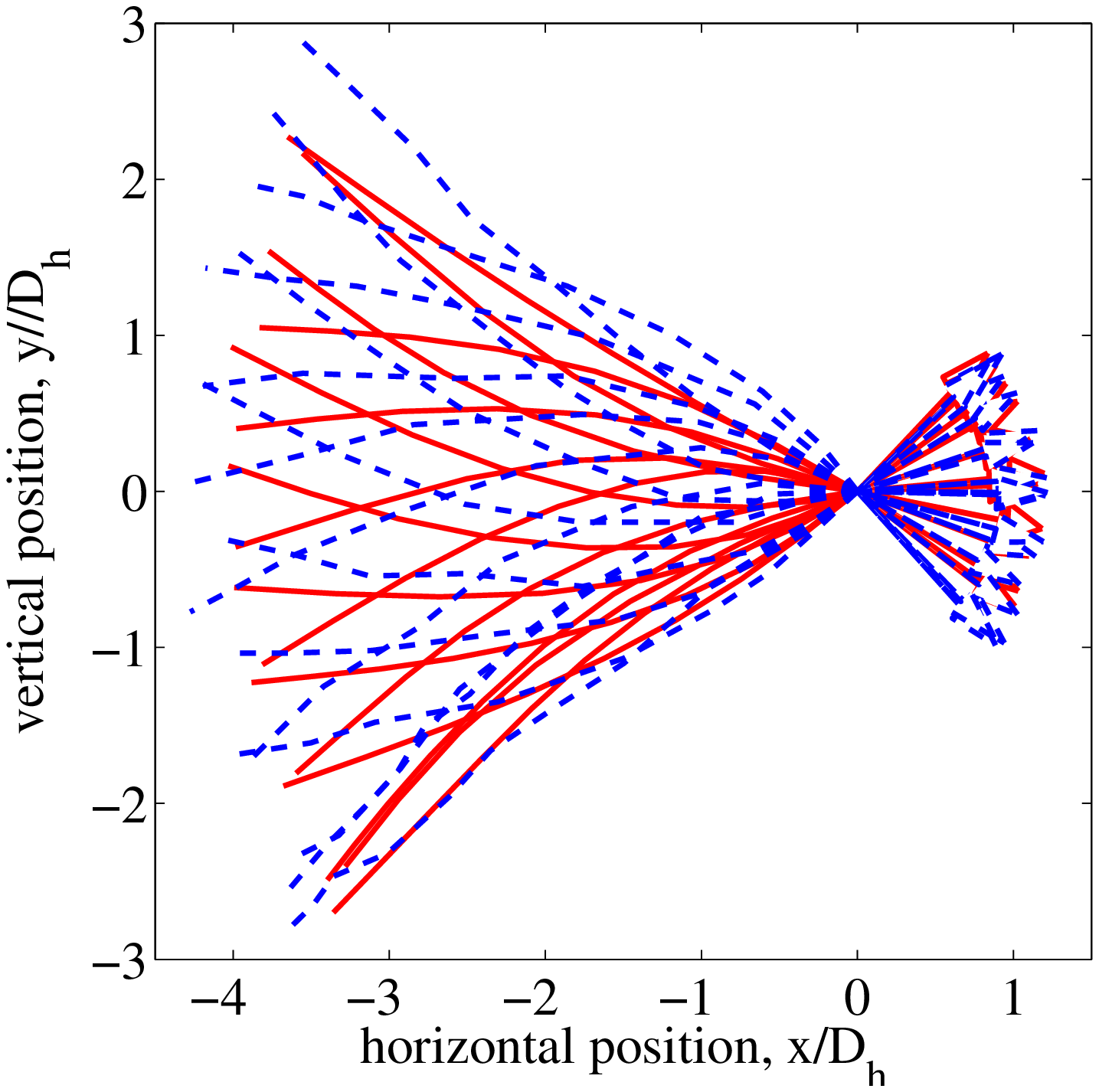}
\caption{(color on line) Superimposed shapes of the flexible tail at different instants throughout the oscillation cycle. The solid (red online) and dashed (blue online) lines display the tail shape for the Newtonian (N2) and Boger (B2) fluids, respectively. These measurements correspond to the data shown in Fig.~\ref{fig:setup2}.}
\label{fig:tail_kin}
\end{figure}

One important aspect that needs to be addressed is the potential difference in tail kinematics between different fluids.  As was pointed out by Teran et.~al \cite{teran2010},  the details of the waveform do  affect the propulsion speed, and swimming enhancement occurs only for specific tail shapes. To address this effect in our experiments, we tracked the shape of the tail
at different instants throughout the oscillation cycle. We show in Fig.~\ref{fig:tail_kin}  the comparison of the tail shape for the same swimmer under identical actuation in fluids N2 (solid line) and B2 (dashed). As was already evident from Fig.~\ref{fig:setup2}, there are some differences. The deformation of the tail in the Boger fluid does appear larger but the tail  curvatures are very similar in both cases. Although we cannot quantify the role that the change in tail kinematics plays in the increase of  the swimming speed, based on the observations Fig.~\ref{fig:tail_kin} we conjecture that the effect is not significant in our experiment, and at least not of the magnitude necessary to obtain the results in  Fig.~\ref{fig:Deborah}.
In order to fully resolve this issue however, experiments would have to be conducted for a swimmer with strictly identical kinematics in all fluids, which we plan to do in the future, enabling to unravel the effects of change in the kinematics vs.~purely elastic contributions.

In summary, we have conducted in this paper an experimental investigation of flexible swimming in non-Newtonian (Boger) fluids. By comparing the locomotion of swimmers with similar mechanical properties and under similar external actuation but in different fluids we demonstrate that swimming is systematically enhanced by fluid elasticity. In contrast with the results of Refs.~\cite{teran2010,Liu2011} showing swimming enhancement only in a range of Deborah numbers, our measurements show that the ratio between the non-Newtonian and Newtonian swimming speeds of flexible swimmers always increase with ${\rm De}$. We conjecture that this enhancement was due to normal stress gradients along the swimming direction.

This research was funded in part by the National Science Foundation (grant CBET-0746285), PAPIIT-UNAM program (IN101312) and the UC MEXUS-CONACYT program.

\bibliographystyle{unsrt}

\begin{thebibliography}{10}
\bibitem{Video2}
Video2. This video shows the swimmer with the rigid tail, in a Newtonian fluid . It was done to confirm that the lack of tail flexibility impedes locomotion. (AIP to insert URL). 
\bibitem{Video3}
Video3. TThis video shows a typical PIV experiment, from which the velocity field around the swimmer was obtained. The case shown corresponds to the swimmer with a flat tail moving in a Boger fluid (B2). (AIP to insert URL). 
\end{thebibliography}

\end{document}